# Room Temperature Magnetocaloric effect in $CrTe_{1-x}Sb_x$ Alloys

M. Kh. Hamad[1*], E. Martinez-Teran[2], Yazan Maswadeh[3], Rami Hamad[1], E.G. Al-Nahari[1], A. A. El-Gendy[2,] and Kh. A. Ziq[1§]

[1]Department of Physics, King Fahd University of Petroleum and Minerals, Dhahran 31261, Saudi Arabia

[2]Department of Physics, University of Texas at El Paso, TX79968, USA

[3]Department of Physics and Science of Advanced Materials Program, Central Michigan University, Mt. Pleasant, Michigan 48859, USA.

**Corresponding authors**: [*] mhamad@kfupm.edu.sa; [§] kaziq@kfupm.edu.sa

We investigate the magnetocaloric effect (MCE), relative cooling power (RCP) and crystalline structure in Sb substituted $CrTe_{1-x}Sb_x$ ($0 \leq x \leq 0.2$) alloy. The Rietveld refinement of the XRD pattern of $CrTe_{1-x}Sb_x$ showed the emerging of pure hexagonal *NiAs* structure with $P6_3/mmc$ (194) space group with increasing Sb substitution. We detect a slight increase in the basal plane *a*-lattice parameter, with a much larger reduction in the *c-axis*. Magnetic isotherms were measured in the temperature range of 50-400K. The results revealed an increase in the maximum entropy change $|S_M(T,H)|$ with Sb-substitutions in the temperature range (~285-325K). Moreover, The RCP values increased by about 33% with 20% Sb substitutions. These findings suggest that $CrTe_{1-x}Sb_x$ alloys can be used in room temperature magnetic cooling at fraction of the coast of pure Gd element the porotype magnetic material for magnetic refrigeration.

## 1- Introduction

Magnetism in Cr-based magnetic materials is receiving increased research interest for possible wide practical applications ranging from Van der Waals ferromagnetic materials, magnetic cooling and various spintronic applications [1-3]. Their electrical transport properties can be tuned using the charge and spin degree of freedom [4, 5]. In particular; reduced dimensional ferromagnetic materials and novel properties are fertile area of research and have great potential in various branches of science: material science, nanotechnology, and chemistry [6-8]. Understanding the nature of the magnetic phase transition in 1 or 2-dimensional system is crucial for realizing several practical applications [2, 9].

Recently, Cr-based ternary chalcogenides ($CrRX_3$) where R is a non-transition metal (Sb, Ge) and X = S, Se, Te have attracted renewed interest for their possible applications in spintronics technologies. These materials exhibit 1, 2 and 3-dimensional crystal structure and anisotropic magnetic properties. The $CrX_6$ octahedral forming infinite, edge-sharing, double rutile chains, and R atoms linking neighboring chains; causing changing angle between double rutile chains [10, 11]. These materials are exfoliatable magnetic layers with van der Waals forces between the magnetic layers [3, 12]. The spins in the individual layers mainly orders ferromagnetically. For example, CrSb is a metallic collinear antiferromagnetic (AFM) with a Néel temperature $T_N$ of about 700 K [13-14]. It consists of Ferromagnetic (FM) planes perpendicular to the c-axis, which are AFM coupled [13]. On the other hand, Chromium telluride (CrTe), a FM material with critical temperature Tc of about 340 K [15-21] with nickel-arsenide (NiAs)-type structures [20]. The reported large variation in Curie temperatures in the literature is probably caused by the often non-stoichiometric formation of $CrTe_x$, which is greatly affected by the preparation method [15, 21], and that the stoichiometric bulk CrTe structure does not exist in pure hexagonal phase at room temperature [22]. The reported magnetic properties [23, 23] of CrTe-CrSb solid solutions, which also crystallize in the hexagonal NiAs structure, indicated that this system would be suitable. The ternary system $CrTe_{1-x}Sb_x$ at different Te and Sb concentrations has also been investigated [13, 23-24] and found to go under ferromagnetic to the antiferromagnetic ground state near the Sb-rich region [20]. Moreover, alloying CrTe with Sb causes a continuous reduction in the ferromagnetic transition. However, for low Sb concentration (~5%) we observed in this paper a slight irregularity in the changes in Tc along with slight enhancement in the saturated magnetization. In general, the variation in Tc were found to agree with de Gennes double exchange in this Ferro-antiferromagnetic system [23, 25]. Sb substitution at the Te site affects the double exchange interaction, which in turns affects the magnetic state and phase transition in the solid solution [25].

In this paper, we investigated the room temperature magnetocaloric effect in $CrTe_{1-x}Sb_x$ magnetic material. The relatively high magnetization along with the increase in the maximum entropy change $|S_M(T,H)|$ with Sb-substitutions in the temperature range (~285-325K) suggests that $CrTe_{1-x}Sb_x$ alloys are good candidate for room temperature magnetocaloric cooling. Understanding the ferromagnetic transition in these materials is a first step towards this goal. In this paper, we investigate the effect of Sb substitution in the *$CrTe_{1-x}Sb_x$* (0.0 ≤ x ≤ 0.20) on the magnetocaloric properties near FM-PM phase transition.

## *2 Experimental Work*

Solid-state reaction has been used to prepare *$CrTe_{1-x}Sb_x$* with *0.0 ≤ x ≤ 0.2* [26]. High purity (4N and 5N) powders of *Cr, Te, and Sb* were used to prepare all the samples. Stoichiometric elements were grinded to a mesh size of 400 or less then pressed in a 5×2 *mm* diameter pellets. The pellets were encapsulated in evacuated quartz tubes partially filled with high purity Argon gas. All sealed samples were gradually heated to 800 ºC and annealed for 10 hours. The samples were re-powdered, pressed then reannealed at 1000 ºC for 24 hours. X-ray diffraction (XRD) analysis was performed on $CrTe_{1-x}Sb_x$ polycrystalline powder using Bruker XRD D2-Phaser with Cu Kα (λ=1.54056 Å) [27]. The XRD-patterns (10-80º range) were analyzed using Rietveld refinements available in FULLPROF software. Magnetic measurements have been carried on a Quantum design 3-Tesla VersaLab in the temperature range 50-400K. A homemade ac-susceptometer has been used to measure the real and imaginary parts of the susceptibility using SR850 Lock-in-amplifier. LABVIEW software was used to control the lock in amplifier and a Lake Shore 336 temperature controller. The amplitude of the excitation ac-field is fixed at 1 $Oe$ and the frequency is 887$Hz$ [28].

*3 Results and discussions:*

*3.1 X-Ray Diffraction*

A selected XRD patterns for $CrTe_{1-x}Sb_x$ (x=0.00, 0.10, and 0.20) along with Rietveld refinement [29, 30] are shown in Fig.1. The Rietveld refinement of XRD pattern of $CrTe_{1-x}Sb_x$ samples revealed a hexagonal structure with P6$_3$/mmc (194) space group in all substituted samples [31]. However, a minority of non-reacted phase in the Sb-free sample (CrTe) has been detected in agreement with Street *et al* results [18]. According to *Street et al*, Chromium telluride of nominal stoichiometry CrTe is existing over a range of stoichiometry with the hexagonal NiAs structure [18, 31]. It is worth mentioning that even after the second prolonged annealing, the pure CrTe sample was found to contain other minor phases. The lattice parameters for all samples $CrTe_{1-x}Sb_x$ with *x ≤ 0.2* are shown in Table 1. The data shows slight increase in the basal plane *a*-lattice parameter. However, these changes are not significant and are consistent with earlier work reported by Lotgering and Gorter on CrTe-CrSb solid solution [16]. A more significant reduction has been observed in the c-lattice parameter. This is qualitatively consistent with the smaller ionic radius of Sb (0.76 Å) compared to Te ionic radius (0.97 Å) [32]. The changes in the Cr-Te/Sb octahedron would appear in the hexagonal unit cell representation, as an atomic bond relaxation over the a-b axes, and relatively more noticeable changes as an atomic stress over the c-axis. Lotgering and Gorter [16] found that the a-axis is almost constant over 25% Sb substitution at the Te site, while the changes in the c-axis are more significantly over the same range. These findings are in line with results reported in Table 1. Moreover, these variations are also reflected in a slight increase in the unit cell volume. These findings further support that Sb is being incorporated in CrTe-structure and that Sb is substituting Te.

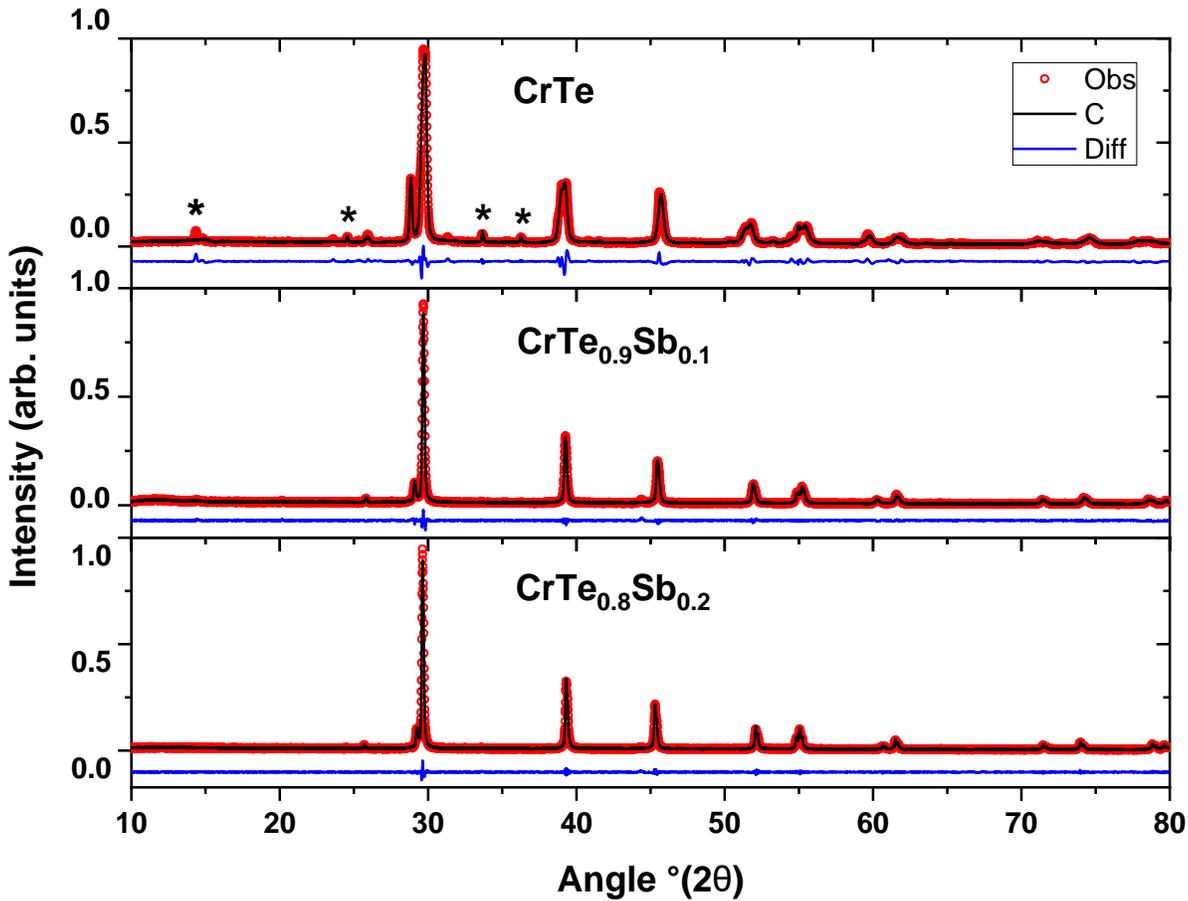

*Figure 1: X-ray diffraction patterns for CrTe1-xSbx (x =0.00, 0.10, and 0.20). The stars indicate the non-reacted impurity phase.*

Table 1: Lattice parameters of $CrTe_{1-x}Sb_x$ ($0.0 \leq x \leq 0.2$) obtained using Rietveld refinement.

| x | a (Å) | c (Å) | c/a | Volume (Å³) |
|---|---|---|---|---|
| 0.00 | 3.969(9) | 6.164(3) | 1.5527 | 84.14 |
|  | 3.982(9) | 6.217(2) | 1.5610 | 85.41 |
| 0.05 | 3.982(4) | 6.177(3) | 1.5511 | 84.48 |
| 0.10 | 3.988(8) | 6.140(7) | 1.5395 | 84.61 |
| 0.15 | 3.999(1) | 6.106(3) | 1.5269 | 84.57 |
| 0.20 | 4.000(0) | 6.103(2) | 1.5258 | 84.57 |

*3.2 Magnetic characterizations*

Variations of the dc-susceptibility was measured in an applied magnetic field of $300 Oe$ in the temperature range of 50- $400K$ for all samples. The thermomagnetic curves *(M vs. T)* of $CrTe_{1-x}Sb_x$ samples (x=0.00, 0.10, and 0.20) are shown in Fig. 2. A clear PM-FM transition can be seen at ~ 325.0, 296.0, and 287.0 K respectively. These transitions are in qualitative agreement with the ac-susceptibility measurements shown in the inset of Fig. 2. Figure 2 also revealed a noticeable increase in the magnetization accompanied with sharper transition with increasing Sb substitution.

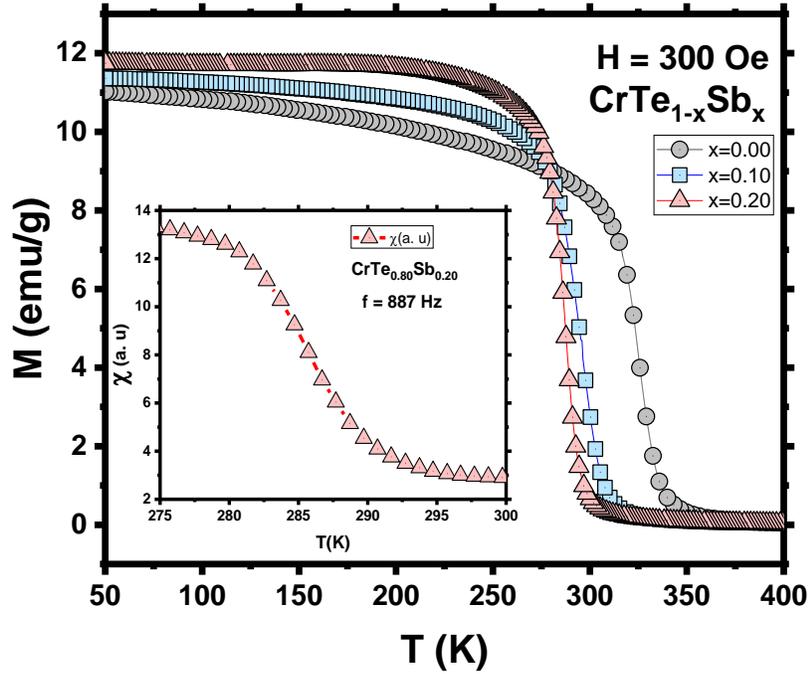

*Figure 2: Temperature dependence of the magnetization in an applied field of 300Oe for CrTe1-xSbc (x=0.0, 0.1, and 0.2). The inset shows the ac-susceptibility for the x=0.2 sample. The amplitude of the excitation ac-field is 1 Oe and the frequency is 887Hz.*

The inverse dc-susceptibility $\chi^{-1}$ at high temperature (250-400K) is shown in Fig. 3 (a). All samples obey Curie-Weiss law in the high temperature region $\chi = C(T - \theta_p)^{-1}$ where C is the Curie constant, and $\theta_p$ is the Curie temperature. A positive $\theta_p$ values reflects ferromagnetic exchange interaction. A Curie-Weiss linear fitting above 350 K yields $\theta_p$ ~ 341.2, 308.8, and 301.6 K for x=0.0, 0.1 and 0.2 respectively. These $\theta_p$ values are significantly higher than the measured Tc, suggesting strong mixed AFM-FM interaction in the samples. Similar behavior has been seen in previous works [3, 10, and 33]. The Tc and $\theta_p$ values for all samples are listed in Table 2.

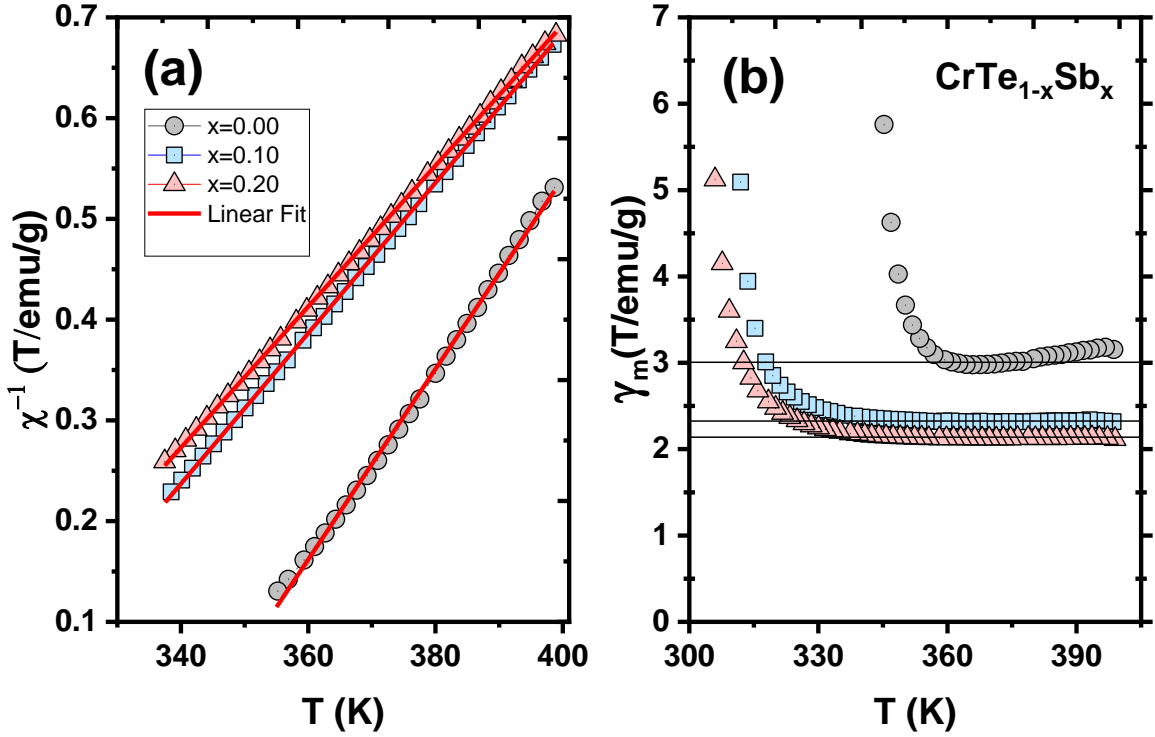

*Figure 3: (a) Temperature-dependent inverse magnetization of $CrTe_{1-x}Sb_x$ (x =0.0, 0.1, and 0.2) measured at 300Oe. (b) Change in the molecular field coefficient with temperature for the samples.*

The effect of the environment can be clearly illustrated using the molecular field theory [34]. The Curie temperature and the magnetic susceptibility are used to calculate the molecular field coefficient $\gamma_m$ using $\chi\gamma_m = \frac{\theta_p}{T-\theta_p}$. The results are shown in Fig. 3 (b). The $\gamma_m$ values for the unsubstituted sample are almost constants down to about 360K. The corresponding values for x=0.10 and x=0.20 are constants down to about 340K. The temperature independent part of $\gamma_m$ for the three samples are also listed in Table 2. Figure 3 (b) reveals a significant reduction in the $\gamma_m$ values with increasing Sb- concentration, this further suggests weakening of the FM molecular field with Sb substitutions.

*Table 2: Critical temperature, Curie temperature, and the molecular field coefficient for the samples used in this study.*

| x | $T_c$ (±0.6) (K) | $\theta_p$ (±0.7) (K) | $\gamma_m$ (±0.2) (T/emu/g) |
|---|---|---|---|
| 0.00 | 325.0 | 342.9 | 3.0 |
| 0.10 | 296.4 | 308.3 | 2.3 |
| 0.20 | 287.2 | 301.0 | 2.1 |

*3.3 Magnetocaloric Effect and Relative Cooling Power:*

The magnetization isotherms ($M$ vs. $H$ curves) for CrTe$_{1-x}$Sb$_x$ (x = 0) near the ferromagnetic transition temperature are presented in Fig. 4(a) with temperature increment $\delta T$=2K. At room temperature, the compound is in its ferromagnetic state, with rapidly increasing magnetization at about 0.2T, reaching saturation near 1T. Moreover, with increasing temperature, the magnetization decreases, and the compound moves toward the paramagnetic (PM) phase at higher temperature. In the PM region, the magnetization is linearly increasing from 0 up to 3T. The magnetization curves of the investigated samples (x=0.0, 0.1 and 0.2) at T=300K are shown in Fig. 4(b). The figure shows that the magnetization is suppressed upon Sb substitution at the same temperature further supporting reducing in the ferromagnetic transition. In addition, the samples with x=0.00 and 0.10 are almost saturated, while the sample with x=0.20 is approaching saturation near 3 Tesla. The figure also shows a gradual reduction in the saturated magnetization with Sb-substitutions. The reduction in the saturated magnetization is an indication of lowering the FM- exchange interaction resulted from Sb-substitutions. The differences in the magnetization of both samples with 0.1 and 0.2 Sb strongly support the difference in Sb concentration in both samples. The variations in the magnetic properties with Sb substitution provides further evidence that 0.1 and 0.2 Sb substituted samples are indeed samples with different Sb-concentrations. Moreover, the samples with Sb=0.1 and 0.2 showed different Curie transition temperature Tc (Fig. 2). These findings qualitatively agree the magnetic measurements and the corresponding phase diagram reported by Takei *et al* [23].

These isotherms are commonly used to investigate several magnetic properties such as the critical behavior and the critical exponents, $\beta, \gamma, and\ \delta$ [3, 12, 35] and the magnetocaloric effect. In the following section, we will present the relative cooling power (RCP), and the change in magnetic entropy $\Delta S_M$ to assess the possibilities of using the CrTe$_{1-x}$Sb$_x$ as magneto-caloric materials at room temperature [8].

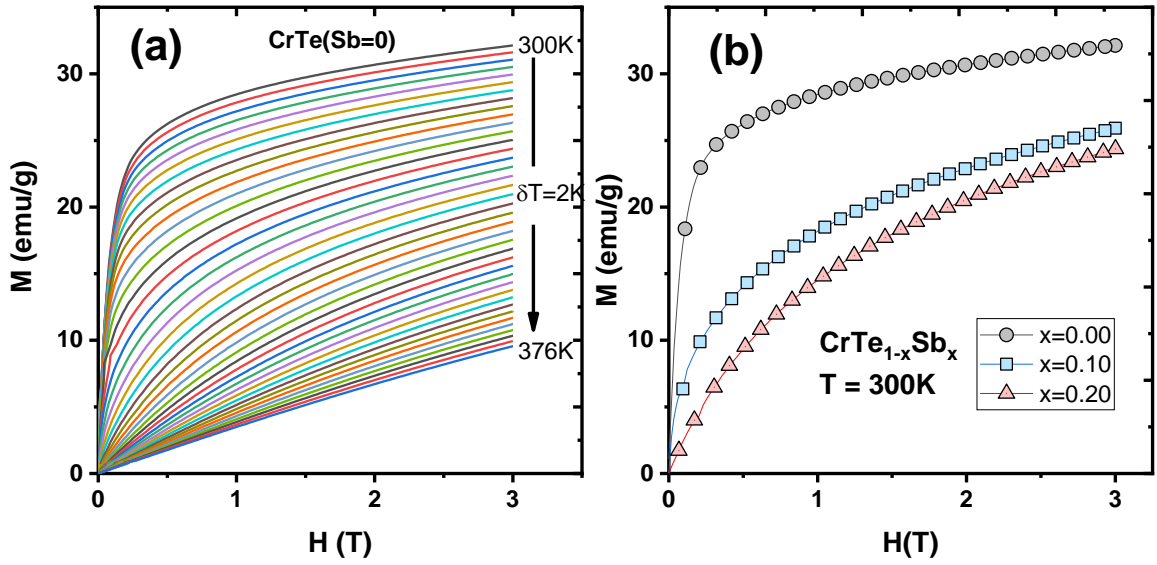

*Figure 4: (a) The isothermal magnetization M(H) at different temperatures for the unsubstituted CrTe sample. (b) Magnetization vs. applied magnetic field at T=300K for the studied samples.*

The change in magnetic entropy $\Delta S_M$ is commonly calculated using the magnetization isotherms (Fig. 4) using Eq. (1).

$$\Delta S_M \left(\frac{T_1+T_2}{2}\right) = \frac{1}{T_1-T_2} \left[\int_0^{H_{max}} M(T_2, H)dH - \int_0^{H_{max}} M(T_1, H)dH\right] \quad (1)$$

Which is approximated at an average field and temperature using Eq. (2)

$$\Delta S_M(T, H) = \sum_i \frac{M_{i+1}(T_{i+1},H) - M_i(T_i,H)}{T_{i+1}-T_i} \Delta H \quad (2)$$

The magnetic entropy depends on the average temperature and the maximum cycling field. The e change in entropy $|\Delta S_M|$ for CrTe$_{1-x}$Sb$_x$ (x=0.00, 0.10, 0.20) samples have been calculated above and below the Ferromagnetic transition temperatures for all samples; the data are shown in Fig. 5. The magnitude of $(-\Delta S_M)$ for all samples showed a maximum at a temperature near the corresponding FM– PM transition temperature. The maximum cycling field is 3Tesla.

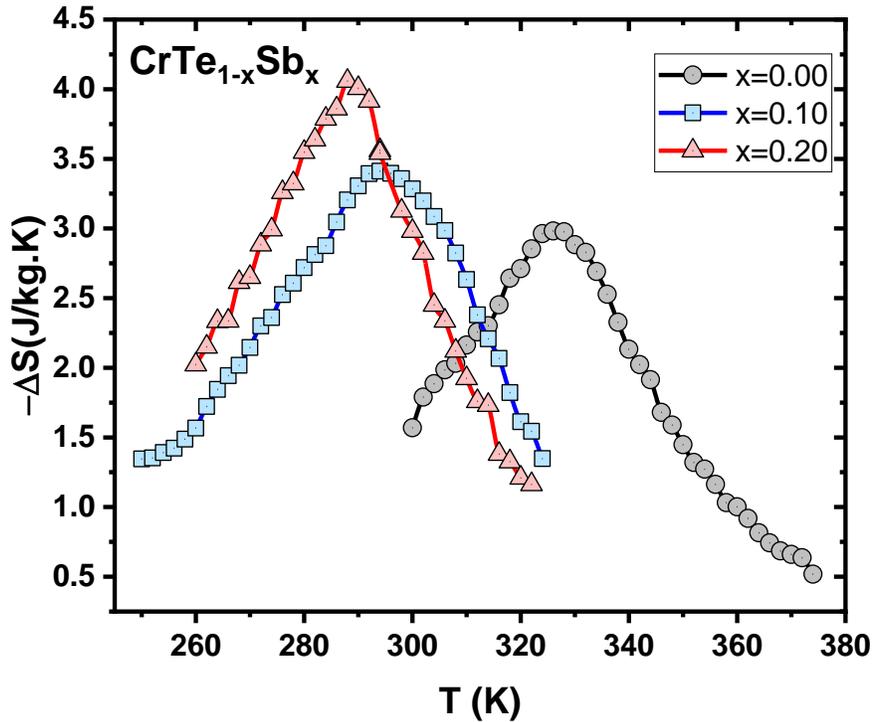

*Figure 5: Temperature dependence of negative magnetic entropy change of CrTe$_{1-x}$Sb$_x$ at maximum cycling field of 3T.*

The variation of the $|\Delta S(T,H)|$ with temperature for cycling field 0-3 Tesla can be used to evaluate the 'Relative Cooling Power (RCP)" introduced by Pecharsky and Gschneidner [36-38]. The RCP is a practical measure to evaluate the refrigeration capacity for magnetocaloric materials. It is, basically, represent the amount of heat that can be transferred from the hot to the cold reservoir.

Moreover, the (RCP) value depends on both, the maximum cycling field, and the maximum value of the magnetic entropy change ($\Delta S_M^{max}$) and the full-width at half-maximum ($\Delta T_{FWHM}$) of the peak. It is given by the following relation [35, 36]:

$$RCP = |\Delta S_M^{max}| \times \Delta T_{FWHM} \qquad (3)$$

The calculated (RCP) values of CrTe$_{1-x}$Sb$_x$ samples obtained at ΔH=3Tesla are presented in Table 3. The data shows a significant increase (~33%) in the RCP values with 20% Sb substitution. We also present in the table reported values of pure gadolinium (Gd) as benchmark for magnetocaloric material obtained at ΔH= 2 and 3 Tesla. A noticeable increase in the FWHM of **ΔS**$_{max}$ is shown in the Table which exceeds the reported values for Gd. The RCP value for CrTe$_{0.80}$Sb$_{0.20}$ is about 61% of RCP value for Gd. The relatively high RCP value and the wide temperature range of $\Delta S_M^{max}$ near room temperature for the investigated samples are characteristics of a promising magnetic refrigerant at ambient temperatures at fraction of the coast of pure Gd.

*Table 3: Maximum entropy change and RCP value for CrTe$_{1-x}$Sb$_x$ samples and Gd for comparison.*

| Material | \|ΔS$_{max}$\| (±0.3) (J/kg.K) | FWHM (±0.5) (K) | ΔH (T) | RCP (±0.8) (J/kg) | Ref. |
|---|---|---|---|---|---|
| Gd | 5.4 | 33 | 2 | 178.5 | [37, 39] |
| Gd | 7.2 | 45 | 3 | 324 | [37] |
| CrTe | 3.0 | 49.3 | 3 | 147.3 | This work |
| CrTe$_{0.90}$Sb$_{0.10}$ | 3.4 | 57.3 | 3 | 195.5 | This work |
| CrTe$_{0.80}$Sb$_{0.20}$ | 4.1 | 48.8 | 3 | 198.3 | This work |

## Conclusions

The Rietveld refinement of XRD pattern of *CrTe$_{1-x}$Sb$_x$* (x=0.0, 0.1 and 0.2) revealed a hexagonal *NiAs* structure with P6$_3$/mmc (194) space group for *0 ≤ x ≤ 0.2*. A minority phase of non-reacted materials was present in the pure stoichiometry *CrTe* sample. All *Sb*-substituted samples have *NiAs* single-phase structure. The data shows a slight increase in the basal plane *a*-lattice parameter and in the unit cell volume with a larger reduction in the *c-axis*. Magnetic isotherms measured near room temperatures, for CrTe$_{1-x}$Sb$_x$ samples, revealed a significant increase in the maximum entropy change |S$_M$(T,H)| with Sb-substitutions. Moreover, The RCP values increased by about 33% with 20% Sb substitutions at the Te-sites. These findings suggest that CrTe$_{1-x}$Sb$_x$ materials can be used in room temperature magnetic refrigeration at fraction of the coast of pure Gd, the prototype magnetic material for magnetic refrigeration.

## Acknowledgment

The authors acknowledge the support provided by the Deanship of Scientific Research at King Fahd University of Petroleum and Minerals (KFUPM) for funding this work under project No. SR191008.